\documentclass[aps,showpacs,twocolumn,superscriptaddress]{revtex4}
\usepackage{graphicx}
\usepackage{dcolumn}

\begin{document}

\title{Faddeev calculation of a $K^- p p$ quasi-bound state}

\author{N.V. Shevchenko\footnote{Corresponding author:
shevchenko@ujf.cas.cz}}
\affiliation{Nuclear Physics Institute, 25068 \v{R}e\v{z}, Czech Republic}

\author{A. Gal}
\affiliation{Racah Institute of Physics, The Hebrew University,
Jerusalem 91904, Israel}

\author{J. Mare\v{s}}
\affiliation{Nuclear Physics Institute, 25068 \v{R}e\v{z}, Czech Republic}

\date{\today}

\begin{abstract}
\rule{0ex}{3ex}

We report on the first genuinely three-body ${\bar K}NN - \pi \Sigma
N$ coupled-channel Faddeev calculation in search for quasi-bound states in
the $K^- p p$ system. The main absorptivity in the $K^- p$ subsystem is
accounted for by fitting to $K^- p$ data near threshold.
Our calculation yields one such quasi-bound state, with
$I=1/2$, $J^{\pi}=0^-$, bound in the range $B \sim 55-70$ MeV,
with a width of $\Gamma \sim 90-110$ MeV. These results differ substantially
from previous estimates, and are at odds with the $K^- p p \to \Lambda p$
signal observed by the FINUDA collaboration.

\end{abstract}

\maketitle


The issue of $\bar K$ nuclear quasi-bound states has attracted
considerable interest recently, motivated by earlier suggestions
for (anti)kaon condensation in dense matter~\cite{BLR94} and by
extrapolations of $K^-$ optical potentials fitted to
$K^-$ atom data~\cite{FGB93,MFG06}. These $K^-$ atom
studies suggested $\bar K$ nuclear potential depths about
$150 - 200$~MeV at nuclear-matter density $\rho_0=0.16~{\rm fm}^{-3}$,
although potentials evaluated by fitting to $K^-p$ low-energy data give
substantially lower values, about 100~MeV~\cite{WKW96} or even as low as
50 MeV~\cite{SKE00} depending on how the $\bar K N$ in-medium
$t$ matrix is constructed. It was pointed out that $\bar K$ nuclear states,
if bound by over 100~MeV where the $\bar K N \to \pi \Sigma$ main
strong-decay channel is closed, might become sufficiently narrow to be
observed~\cite{Wyc86,Kis99,AYa99}. Yamazaki and Akaishi~\cite{YAk02},
in particular, discussed few-body $\bar K$ nuclear configurations
in which the strongly attractive $I=0~\bar K N$ interaction is maximized.
It is the $I=0$ coupled-channel $s$-wave interaction that generates
a resonance in the $\pi \Sigma$ coupled channel about 27 MeV below
the $K^-p$ threshold, the quasi-bound $\Lambda(1405)$~\cite{DDe91}.
The lightest $\bar K$ {\it nuclear} configuration maximizing the
$I=0~\bar K N$ interaction is the $I=1/2$ $\bar K (NN)_{I=1}$
state with $S=L=0$ and $J^{\pi}=0^-$~\cite{Nog63}.
The significance of identifying this potentially low-lying quasi-bound state
in the $K^-pp$ mass spectrum of suitably chosen production
reactions has been recently emphasized~\cite{YAk06}.
However, because the coupling of the two-body $K^-p$ channel to the
absorptive $\pi Y$ channels was substituted by an energy-independent
complex $\bar K N$ potential, the results for binding energy and width
of the $K^-pp$ system~\cite{YAk02} provide at best only a rough estimate.
Recently, the FINUDA collaboration at DA$\Phi$NE,
Frascati, presented evidence in $K^-$ stopped reactions on several nuclear
targets for the process $K^-pp \to \Lambda p$, interpreting the observed
signal as due to a $K^-pp$ deeply bound state~\cite{ABB05}. However, this
interpretation has been challenged in Refs.~\cite{MFG06,MOR06}.
Given this unsettled experimental search for a quasi-bound $K^-pp$ state,
precise three-body calculations for the $K^-pp$ system appear well motivated
at present.

In this Letter we report on the first $\bar K NN - \pi \Sigma N$
coupled-channel Faddeev calculation which is genuinely three-body calculation,
searching for quasi-bound states that are experimentally accessible through
a $K^-pp$ final state. Coupled-channel three-body Faddeev calculations
were reported for $K^-d$, with an emphasis on other
entities than on quasi-bound states~\cite{TGE79}. We note that the $K^-d$
system is not as favorable as the $K^-pp$ system for strong binding, since
the relative weight of the $I=0~\bar K N$ interaction with respect to the
weakly attractive $I=1~\bar K N$ interaction is $1:3$ for $K^-d$ and
$3:1$ for $(K^-pp)_{I=1/2}$. By doing coupled channel calculations, with
two-body input fitted to available low-energy data, we wish to determine
the scale of binding energy and width expected for few-body $\bar K$ nuclear
systems.



In the present work we solve non-relativistic three-body Faddeev
equations in momentum space, using the
Alt-Grassberger-Sandhas (AGS) form~\cite{AGS67}. The AGS equations
for three particles are:
\begin{eqnarray}\nonumber
U_{11} &=& \phantom{G_0^{-1}  +} T_2 G_0 U_{21} + T_3 G_0 U_{31} \\
U_{21} &=& G_0^{-1} + T_1 G_0 U_{11} + T_3 G_0 U_{31} \\ \nonumber
U_{31} &=& G_0^{-1} + T_1 G_0 U_{11} + T_2 G_0 U_{21} ~,
\end{eqnarray}
where $G_0$ is the free three-body Green's function and $T_i,~i=1,2,3,$ are
two-body T matrices in the three-body space for the pair excluding particle
$i$. These equations define three unknown transition operators $U_{ij}$
describing the elastic and re-arrangement processes:
\begin{eqnarray}\nonumber
U_{11}: \qquad 1 + (23) & \to& 1 + (23) \\
U_{21}: \qquad 1 + (23) & \to& 2 + (31) \\ \nonumber
U_{31}: \qquad 1 + (23) & \to& 3 + (12) ~,
\end{eqnarray}
with Faddeev indices $i, j = 1,2,3$ denoting simultaneously a given
particle and its complementary interacting pair. Since the ${\bar K}N$
two-body subsystem is strongly coupled to other channels, particularly
via the $\Lambda(1405)$ resonance to the $I=0$ $\pi \Sigma$ channel,
it is necessary to extend the AGS formalism in order to include these
channels explicitly. Thus, all operators entering the AGS equations
become $3 \times 3$ matrices: $G_0 \to G_0^{\alpha \beta}
= \delta_{\alpha \beta} G_0^{\alpha}$ which is diagonal in the channel
space, and $T_i \to T_i^{\alpha \beta}$ where $\alpha, \beta$ are channel
indices as follows:
\begin{equation}
\label{eq:T}
 T_1 =  \left(
    \begin{tabular}{ccc}
    $T_1^{NN}$ & 0 & 0 \\
    0 & $T_1^{\Sigma N}$ & 0 \\
    0 & 0 & $T_1^{\Sigma N}$
    \end{tabular}
 \right)
\end{equation}
$$
 T_2 = \left(
    \begin{tabular}{ccc}
    $T_2^{KK}$ & 0 & $T_2^{K \pi}$ \\
    0 & $T_2^{\pi N}$ & 0 \\
    $T_2^{\pi K}$ & 0 & $T_2^{\pi \pi}$
    \end{tabular}
 \right) \,
 T_3 =  \left(
    \begin{tabular}{ccc}
    $T_3^{KK}$ & $T_3^{K \pi}$ & 0 \\
    $T_3^{\pi K}$ & $T_3^{\pi \pi}$ & 0 \\
    0 & 0 & $T_3^{\pi N}$
    \end{tabular}
 \right).
$$
We assign particle labels ($1,2,3$) to
($\bar{K},N,N$) in channel~1, to ($\pi, \Sigma, N$) in channel~2 and
to ($\pi, N, \Sigma$) in channel~3. Here $T^{NN}$,
$T^{\pi N}$ and $T^{\Sigma N}$ are one-channel $T$-matrices,
whereas $T^{KK}$, $T^{\pi\pi}$, $T^{\pi K}$ and $T^{K \pi}$ are the elements
of the two-channel $T^{KN - \pi \Sigma}$ matrix, accounting for $\bar K N
\to \bar K N$ and $\pi \Sigma \to \pi \Sigma$ elastic processes,
and for $\bar K N \to \pi \Sigma$ and $\pi \Sigma \to \bar K N$
inelastic transitions, respectively. We neglect the $I=1$ inelastic
transition $\bar K N \to \pi \Lambda$ since experimentally it is
outweighed by the $\bar K N \to \pi \Sigma$ transition, and also since the
$I=1$ $\bar K N$ configuration plays a minor role in the structure of the
$I=1/2$ $K^- pp$ system under discussion.
Upon this extension into channel space, the unknown operators $U$
assume the most general matrix form:
$U_{ij} \to U_{ij}^{\alpha \beta}$. Substituting these new $3 \times 3$
operators into the AGS system of equations we obtain the system to
be solved.

Assuming charge independence, three-body quasi-bound states are labelled
by isospin. The isospin basis is used throughout
our calculation within a coupling scheme that ensures that we are searching
for an $I=1/2$ quasi-bound state. Assuming pairwise $s$-wave meson-baryon
interactions, as appropriate to the $\bar K N - \pi \Sigma$ system near
the $\bar K NN$ threshold, and $s$-wave baryon-baryon interactions limited
to the $^1S_0$ configuration as appropriate to $pp$, the total spin and
total orbital angular momentum of the three-body system are $S=L=0$.
Tensor forces are not operative for this situation, which also reinforces
the neglect of coupling to $\pi \Lambda N$ since the strong
$\Sigma N \to \Lambda N$ transition is dominated by the tensor force in the
$^3S_1$ $YN$ configuration. Hence, the $\bar{K}NN - \pi \Sigma N$ system
explored in this Faddeev calculation has quantum numbers
$I=1/2,~L=0,~S=0,~J^{\pi}=0^-$.

In order to reduce the dimension of the integral equations,
a separable approximation for the two-body $T$ matrices is used:
\begin{equation}
\label{eq:Toperator}
 T_{i,I_i}^{\alpha \beta} = |g_{i,I_i}^{\alpha} \rangle
\tau_{i,I_i}^{\alpha \beta} \langle g_{i,I_i}^{\beta} | ~,
\end{equation}
where $I_i$ is the conserved isospin of the interacting pair.
[for $\alpha = \beta$ our generalized $T$-matrices coincide with the usual
ones.] For separable two-body $T$-matrices, the AGS equations may be rewritten
using a new kernel and unknown functions:
\begin{eqnarray}
Z_{ij, I_i I_j}^{\alpha \beta} &\equiv&  \delta_{\alpha \beta} \,
\langle g_{i,I_i}^{\alpha} | G_0^{\alpha} | g_{j,I_j}^{\beta}
\rangle   \\
X_{ij, I_i I_j}^{\alpha \beta} &\equiv&  \langle
g_{i,I_i}^{\alpha} | G_0^{\alpha} \, U_{ij, I_i I_j}^{\alpha
\beta} \, G_0^{\beta} | g_{j,I_j}^{\beta} \rangle ~,
\end{eqnarray}
respectively. The calculation of the kernels $Z$ involves
transformation from  one set of Jacobi coordinates to another one
and isospin recoupling as well.
The position of the three-body pole was searched as a zero of the determinant
of the kernel of the system of integral equations on the corresponding
unphysical sheet.
More details on the extended AGS equations and
the numerical procedure are relegated to an expanded
version of this paper. Here it suffices to mention that by assuming
charge independence, $s$-wave pairwise interactions,
and antisymmetrizing over the two nucleons, we end up in a system
of nine coupled integral equations. This is the {\it minimal}
dimensionality of any Faddeev calculation in the $I=1/2~J^{\pi}=0^-$
sector which attempts to account explicitly for the strong absorptivity
of the $\bar K N$ interactions near threshold. Within this scheme, the
interaction of the relatively energetic pion with the slow baryons was
neglected, partly because its $p$-wave nature would require an
extension of the present $s$-wave calculation.


\begin{table}
\caption{Strength parameters $\lambda_{I=0,1}^{\alpha \beta}$
(in units fm$^{-2}$) for the $\bar{K}N -\pi \Sigma$ potentials
(\ref{eq:Vseprb}) with range parameter
$\beta = 3.5~{\rm fm}^{-1}$, corresponding to
$a_{K^-p}=(-0.70+{\rm i}~0.60)~{\rm fm}$.}
\label{smg06.tabl1}
\begin{tabular}{cccccc}
\hline \hline \noalign{\smallskip}
$\lambda_{I=0}^{\bar{K}N,\bar{K}N}$ & $\lambda_{I=0}^{\bar{K}N,\pi \Sigma}$ &
$\lambda_{I=0}^{\pi \Sigma,\pi \Sigma}$ & $\lambda_{I=1}^{\bar{K}N,\bar{K}N}$ &
$\lambda_{I=1}^{\bar{K}N,\pi \Sigma}$ & $\lambda_{I=1}^{\pi \Sigma,\pi \Sigma}$
\\ \noalign{\smallskip} \hline \noalign{\smallskip}
 -1.370 & 1.414 & -0.176 & 0.007 & 1.734 & -0.340 \\
\hline \hline
\end{tabular}
\end{table}

The input separable potentials for the $T$-matrices (\ref{eq:Toperator})
are given in momentum space by
\begin{equation}
\label{eq:Vseprb}
 V_{I}^{\alpha \beta}(k_{\alpha},k'_{\beta}) = \lambda_{I}^{\alpha \beta} \,
 g_{I}^{\alpha}(k_{\alpha}) \, g_{I}^{\beta}(k'_{\beta}) \, ,
\end{equation}
where $k_{\alpha}, k'_{\beta}$ are two-particle relative momenta in the
two-body respective channels, and $\lambda_{I}^{\alpha \beta}$ are
strength-parameter constants.
For the $\alpha = \beta = (NN)_{I=1}$ channel,
we have used a separable approximation
of the Paris potential~\cite{ZPH83}, corresponding to the one-rank
potential~(\ref{eq:Vseprb}) with $\lambda_{I}^{NN} = -1$ and a form factor:
\begin{equation}
\label{eq:Paris}
g_{I=1}^{NN}(k) = \frac{1}{2 \sqrt{\pi}} \, \sum_{i=1}^6
\frac{c_{i,I=1}^{NN}}{k^2 + (\beta_{i,I=1}^{NN})^2} \,\, .
\end{equation}
The constants $c_{i,I=1}^{NN}$ and $\beta_{i,I=1}^{NN}$
are listed in Ref.~\cite{ZPH83}.

For the $S=-1$ interactions, the form factors $g_{I}^{\alpha}(k_{\alpha})$
in Eq.~(\ref{eq:Vseprb}) were parameterized by a Yamaguchi form
\begin{equation}
g_{I}^{\alpha}(k^{\alpha}) = \frac{1}{(k^{\alpha})^2 +
(\beta_{I}^{\alpha})^2} \,\, .
\label{eq:Yamaguchi}
\end{equation}
For the $I=3/2$ $\Sigma N$ interaction we made two different choices of
$\lambda_{I=3/2}^{\Sigma N}$
and $\beta_{I=3/2}^{\Sigma N}$. The first choice, labelled (i) below,
reproduces the scattering length~$a_{I=3/2} = 3.8$~fm and effective
range~$r_{I=3/2} = 4.0$~fm of the Nijmegen Model F~\cite{NRS79}.
The second choice, labelled (ii) below, reproduces the most recent
Nijmegen $YN$ phase shifts~\cite{Nij06} using a scattering length
$a_{I=3/2} = 4.15$~fm and effective range $r_{I=3/2} = 2.4$~fm.
For the $I=1/2$ $\Sigma N$ interaction we reproduced the value quoted
by Dalitz~\cite{Dal81} for the scattering length $a_{I=1/2} = -0.5$~fm.

\begin{figure}[t]
\centerline{\includegraphics[scale=0.35,angle=90]{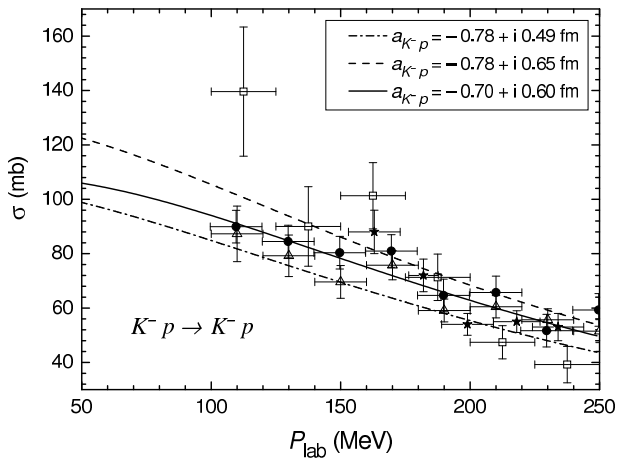}}
\caption{\label{smg06.fig1} Calculated $K^- p \to K^- p$ cross
sections, for three different sets of $\bar K N - \pi \Sigma$
parameters, in comparison with the measured cross sections (see
text).}
\end{figure}

\begin{figure}
\centerline{\includegraphics[scale=0.35,angle=90]{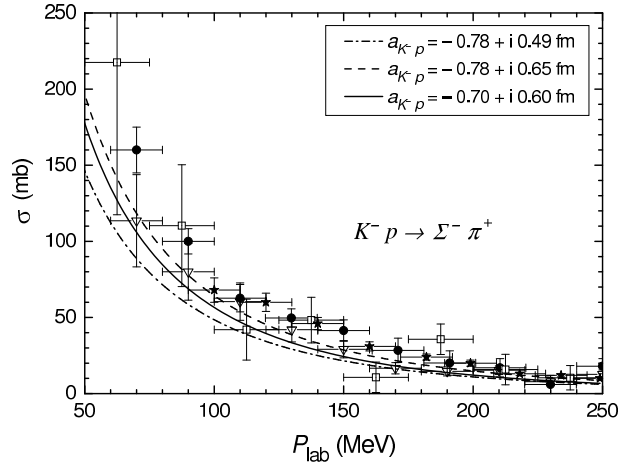}}
\caption{\label{smg06.fig2} Calculated $K^- p \to \pi^+ \Sigma^-$
cross sections, for three different sets of $\bar K N - \pi \Sigma$
parameters, in comparison with the measured cross sections (see text).}
\end{figure}

For the $I=0,1$ $\bar{K}N -\pi \Sigma$ coupled-channel potentials,
the parameters $\lambda_{I=0,1}^{\alpha \beta}$ and $\beta_{I=0,1}^{\alpha}$
in Eqs.~(\ref{eq:Vseprb},~\ref{eq:Yamaguchi}) were fitted to reproduce
(i) $E_{\Lambda(1405)} = 1406.5 - {\rm i}~25$ MeV~\cite{DDe91},
the position and width of $\Lambda(1405)$ which is assumed to be
a quasi-bound state in the $\bar{K}N$ channel and a resonance in the
$\pi \Sigma$ channel, (ii) the branching ratio at rest~\cite{TDS71}
$
\gamma = {\Gamma(K^- p \to \pi^+ \Sigma^-)}/{\Gamma(K^- p \to
\pi^- \Sigma^+)} = 2.36 \,,
$
and (iii) the $K^- p$ scattering length $a_{K^-p}$ for which we used as a
guideline the KEK measured value~\cite{IHI97}:
\begin{equation}
a_{K^-p}=(-0.78 \pm 0.15 \pm 0.03)+{\rm i}(0.49 \pm 0.25 \pm 0.12)~{\rm fm}
\,.
\label{eq:KEK}
\end{equation}
In order to check the sensitivity of our results to this input,
within the quoted errors, we fitted three different values of $a_{K^-p}$
using a range parameter $\beta = 3.5~{\rm fm}^{-1}$. All three sets of our
$\bar K N - \pi \Sigma$ parameters, which also reproduce the energy and
width of $\Lambda(1405)$ and the branching-ratio $\gamma$,
yield low-energy $K^- p \to K^- p$ and $K^- p \to \pi^+ \Sigma^-$
cross-sections which are in a good agreement with experimental
data, as shown in Figs.~\ref{smg06.fig1} and~\ref{smg06.fig2}.
We note that the data points in these figures are precisely those
compiled and cited in Ref.~\cite{BNW05}.
The strength parameters $\lambda_{I=0,1}^{\alpha \beta}$ for the
$\bar{K}N -\pi \Sigma$ coupled-channel separable potentials
fitted to $a_{K^-p}=(-0.70+{\rm i}~0.60)~{\rm fm}$
are given for illustration in Table~\ref{smg06.tabl1}.



In a test calculation we first switched off the coupling of the $\bar{K}NN$
channels to the $\pi \Sigma N$ channels. This reduces the number of
coupled integral equations from nine to three within the three-body
$\bar{K}NN$ space. We assumed the $\Lambda(1405)$ to be a genuine bound
state of the $(\bar{K}N)_{I=0}$ subsystem, reproducing the real part of the
$K^- p$ scattering length of Eq.~(\ref{eq:KEK}). We found a zero-width bound
state at energy $E_{\bar K NN} = -43.7$~MeV below the $\bar{K}NN$ threshold.
This binding energy is considerably larger than the value
$E_{\bar K NN} \approx -10$~MeV estimated by Nogami~\cite{Nog63}.
We then performed full $\bar{K}NN - \pi \Sigma N$ three-body calculations
for the three sets of $\bar K N - \pi \Sigma$ parameters and for the two
sets (i) and (ii) of $(\Sigma N)_{I=3/2}$ parameters described above.
The sensitivity of the results to the $\Sigma N$ interaction was also
studied by setting $T^{\Sigma N}=0$ for both $I=1/2$ and $I=3/2$.
The calculated binding energies ($B=-E_B$) and widths ($\Gamma$) are presented
in Table~\ref{smg06.tabl2} where the energies are given with respect to the
$K^-pp$ threshold. It is seen that the $\Sigma N$ interaction, dominantly
in the $I=3/2$ channel, adds only about 3 MeV to the binding energy
(less than $6\%$) affecting the width by up to 2 MeV (less than $2\%$).
This is negligible on the scale of binding energies and widths displayed
in the table and is consistent with the negligible effect (less than $2\%$)
that the $YN$ and $\pi N$ final-state interactions were found to have in
the latest $K^-d$ Faddeev calculation of Ref.~\cite{TGE79}.
In contrast, the calculated binding energies and widths show
sensitivity to the fitted $\bar{K} N - \pi \Sigma$ coupled-channel two-body
interactions, giving rise in our calculations to up to about $25\%$ variation
in $B$ and up to about $15\%$ variation in $\Gamma$. It is worth noting that
$B$ increases with ${\rm Im}~a_{K^-p}$, whereas $\Gamma$ is correlated more
with ${\rm Re}~a_{K^-p}$; this feature is typical to strong-absorption
phenomena where the width gets saturated beyond a critical value of
absorptivity~\cite{FGa99}. We have also studied the dependence of the
calculated binding energy and width on the range parameter $\beta$
within acceptable fits, keeping $a_{K^-p}$ constant,
say $a_{K^-p}=(-0.78+{\rm i}~0.49)~{\rm fm}$. The binding energy changes very
little, by about 3 MeV, whereas the width changes appreciably, decreasing
from 115 MeV for $\beta = 3~{\rm fm}^{-1}$ to 89 MeV for $\beta = 4~{\rm fm}^{-1}$.

\begin{table}
\caption{Calculated energy $E_{\bar K NN} = E_B - {\rm i}~ \Gamma /2$
(in MeV) of the $I=1/2,~J^{\pi}=0^-$ quasi-bound $\bar{K} (NN)_{I=1}$
state with respect to the $K^-pp$ threshold, calculated for different
two-body input. $E^{({\rm i})}$ and $E^{({\rm ii})}$ correspond to sets
(i) and (ii), respectively, of the $(\Sigma N)_{I=3/2}$ interaction
parameters, whereas $E^{(0)}$ stands for no $\Sigma N$ interaction
(see text).}
\label{smg06.tabl2}
\begin{tabular}{cccc}
\hline \hline \noalign{\smallskip}
$a_{K^-p}$~(fm) & $E_{\bar K NN}^{({\rm i})}$~(MeV) &
$E_{\bar K NN}^{({\rm ii})}$~(MeV) & $E_{\bar K NN}^{(0)}$~(MeV)  \\
\noalign{\smallskip} \hline \noalign{\smallskip}
$-0.78+{\rm i}~0.49$ & $-55.8-{\rm i}~49.1$ & $-56.2-{\rm i}~50.1$ &
$-53.4-{\rm i}~49.2$ \\
$-0.78+{\rm i}~0.65$ & $-69.4-{\rm i}~46.8$ & $-70.0-{\rm i}~47.9$ &
$-66.3-{\rm i}~47.5$ \\
$-0.70+{\rm i}~0.60$ & $-66.0-{\rm i}~54.7$ & $-66.5-{\rm i}~55.8$ &
$-63.5-{\rm i}~54.6$ \\
\hline \hline
\end{tabular}
\end{table}

Our calculations confirm the existence of an $I=1/2$, $J^{\pi}=0^-$ three-body
quasi-bound state, with appreciable width, in the $\bar K (NN)_{I=1}$ channel.
The width of this quasi-bound state is a measure of its coupling to the
$\pi \Sigma N$ channels where it shows up as a broad resonance.
The coupling to the $\pi \Sigma N$ channels, in addition to providing a width
which renders the $\bar K (NN)_{I=1}$ bound state into a quasi-bound state,
also provides substantial extra attraction through which the binding energy
is increased from 44 MeV to the range of values shown in the table.
The acceptable parameter sets considered in our calculations yield binding
in the range $B \sim 55-70$~MeV, with a width of $\Gamma \sim 90-110$~MeV.
Although the binding energy calculated here is similar to that estimated
by Yamazaki and Akaishi~\cite{YAk02} for $K^-pp$, our calculated width
is considerably larger than their estimate $\Gamma = 61$~MeV  and is also
larger than the width $\Gamma \approx 67$~MeV of the $K^-pp \to \Lambda p$
signal in the FINUDA experiment~\cite{ABB05}. Our range of calculated
binding energies is considerably lower than $B \approx 115$~MeV attributed
by the FINUDA collaboration to a $K^-pp$ bound state. Possible extensions
of the present coupled-channel Faddeev calculation should include the
$I=1~\pi \Lambda$ channel, enlarge the model space to include $p$-wave
two-body interactions and introduce relativistic kinematics.
Relying on the experience of coupled-channel Faddeev calculations of the
$K^-d$ system~\cite{TGE79}, none of these extensions is expected to change
{\it qualitatively} our results and conclusions.

In conclusion, we performed the first coupled-channel three-body Faddeev
calculation for the $I=1/2$ $\bar K (NN)_{I=1}$ system in search of
a quasi-bound state. This state can be reached in production reactions
aiming at a final $K^-pp$ system. It is primarily the large width, here
calculated for a $\bar K$ nuclear state above the $\pi \Sigma$ two-body
threshold, that poses a major obstacle to observing and identifying $\bar K$
nuclear quasi-bound states. Yet, even for deeper states below the $\pi \Sigma$
threshold, in heavier nuclei, a residual width of order 50 MeV is expected to
persist due to $\bar K NN \to YN$ absorption~\cite{MFG06}.


This work was supported by the GA AVCR grant A100480617 and by the Israel
Science Foundation grant 757/05. NVS is grateful to J\'{a}nos R\'{e}vai
for many fruitful discussions. AG acknowledges the support of the Alexander
von Humboldt Foundation and thanks Wolfram Weise for his kind hospitality
at TU Muenchen and for stimulating discussions.



\begin{thebibliography}{99}

\bibitem{BLR94} G.E. Brown, C.-H. Lee, M. Rho, V. Thorsson, Nucl. Phys. A 567
(1994) 937.

\bibitem{FGB93} E. Friedman, A. Gal, C.J. Batty, Phys. Lett. B 308 (1993) 6;
Nucl. Phys. A 579 (1994) 518; E. Friedman, A. Gal, J.~Mare\v{s}, A. Ciepl\'y,
Phys. Rev. C 60 (1999) 024314.

\bibitem{MFG06} J. Mare\v{s}, E. Friedman, A. Gal, Phys. Lett. B 606 (2005)
295; Nucl. Phys. A 770 (2006) 84.

\bibitem{WKW96} T. Waas, N. Kaiser, W. Weise, Phys. Lett. B 379 (1996) 34.

\bibitem{SKE00} J. Schaffner-Bielich, V. Koch, M. Effenberger, Nucl. Phys. A
669 (2000) 153; A. Ramos, E. Oset, Nucl. Phys. A 671 (2000) 481; A. Ciepl\'y,
E. Friedman, A. Gal, J.~Mare\v{s}, Nucl. Phys. A 696 (2001) 173.

\bibitem{Wyc86} S. Wycech, Nucl. Phys. A 450 (1986) 399c.

\bibitem{Kis99} T. Kishimoto, Phys. Rev. Lett. 83 (1999) 4701.

\bibitem{AYa99} Y. Akaishi, T. Yamazaki, in {\it Proc. DA$\Phi$NE Workshop},
Frascati Physics Series XVI (1999) 59; Phys. Rev. C 65 (2002) 044005.

\bibitem{YAk02} T. Yamazaki, Y. Akaishi, Phys. Lett. B 535 (2002) 70.

\bibitem{DDe91} R.H. Dalitz, A. Deloff, J. Phys. G 17 (1991) 289; see also the
latest PDG Tables at J. Phys. G 33 (2006) 1.

\bibitem{Nog63} Y. Nogami, Phys. Lett. 7 (1963) 288.

\bibitem{YAk06} T. Yamazaki, Y. Akaishi, arXiv:nucl-th/0604049.

\bibitem{ABB05} M. Agnello {\it et al.}, Phys. Rev. Lett. 94 (2005) 212303.

\bibitem{MOR06} V.K. Magas, E. Oset, A. Ramos, H. Toki, Phys. Rev. C 74
(2006) 025206.

\bibitem{TGE79} G. Toker, A. Gal, J.M. Eisenberg,
Nucl. Phys. A 362 (1981) 405; M. Torres, R.H.~Dalitz, A. Deloff,
Phys. Lett. B 174 (1986) 213;
A. Bahaoui, C. Fayard, T.~Mizutani, B. Saghai,
Phys. Rev. C 68 (2003) 064001.

\bibitem{AGS67} E.O. Alt, P. Grassberger, W. Sandhas, Nucl. Phys. B 2
(1967) 167.

\bibitem{ZPH83} H. Zankel, W. Plessas, J. Haidenbauer, Phys. Rev. C 28 (1983)
538.

\bibitem{NRS79} M.M. Nagels, T.A. Rijken, J.J. de Swart, Phys. Rev. D 20
(1979) 1633.

\bibitem{Nij06} http://nn-online.org by the Nijmegen group.

\bibitem{Dal81} R.H. Dalitz, Nucl. Phys. A 354 (1981) 101c.

\bibitem{TDS71} D.N. Tovee {\it et al.}, Nucl. Phys. B 33 (1971) 493;
R.J.~Nowak {\it et al.}, Nucl. Phys. B 139 (1978) 61.

\bibitem{IHI97} M. Iwasaki {\it et al.}, Phys. Rev. Lett. 78 (1997) 3067;
T.M.~Ito {\it et al.}, Phys. Rev. C 58 (1998) 2366. We did not use the more
recent DEAR value~\cite{BBC05} since it cannot be reconciled with the other
low-energy $K^-p$ data~\cite{BNW05}.

\bibitem{BBC05} G. Beer {\it et al.}, Phys. Rev. Lett. 94 (2005) 212302.

\bibitem{BNW05} B. Borasoy. R. Ni{\ss}ler, W. Weise, Phys. Rev. Lett. 94
(2005) 213401; Eur. Phys. J. A 25 (2005) 79.

\bibitem{FGa99} E. Friedman, A. Gal, Phys. Lett. B 459 (1999) 43;
Nucl. Phys. A 658 (1999) 345.

\end{thebibliography}
\end{document}